\documentclass[usletter,conference,hyphens,twocolmun]{IEEEtran}
\usepackage[switch,pagewise]{lineno}
\usepackage[utf8]{inputenc}
\usepackage{threeparttable}
\usepackage{cite}
\usepackage{multirow}
\usepackage{graphicx}
\usepackage{hyperref}
\usepackage{url} 
\usepackage{doi}
\usepackage{siunitx}
\usepackage[cmex10]{amsmath}
\usepackage{dblfloatfix}
\usepackage{here}
\usepackage{bm}
\usepackage{listings}
\usepackage[usenames]{color}
\usepackage{algpseudocode,algorithm}
\usepackage{etoolbox}

\usepackage{tikz}
\usepackage{textcomp}

\nolinenumbers
\newcommand{\algcapsize}{\footnotesize}

\makeatletter
\renewcommand{\ALG@beginalgorithmic}{\footnotesize}
\makeatother

\floatname{algorithm}{\algcapsize Algorithm}

\makeatletter
\newcommand{\removelatexerror}{\let\@latex@error\@gobble}
\newcommand{\SP}{\scalebox{0.5}{($\mathrm{SP}$)}}
\makeatother

\renewrobustcmd{\bfseries}{\fontseries{b}\selectfont}
\def\lsim{\mathrel{\rlap{\lower4pt\hbox{\hskip1pt$\sim$}}
    \raise1pt\hbox{$<$}}}                

\IEEEoverridecommandlockouts

\newcommand\copyrighttext{%
  \footnotesize \textcopyright 
  2017 IEEE. Personal use of this material is permitted. 
  Permission from IEEE must be obtained for all other uses, in any current or future media, 
  including reprinting/republishing this material for advertising or promotional purposes, 
  creating new collective works, for resale or redistribution to servers or lists, 
  or reuse of any copyrighted component of this work in other works.}
\newcommand\copyrightnotice{%
\begin{tikzpicture}[remember picture,overlay]
\node[anchor=south,yshift=10pt] at (current page.south) {\fbox{\parbox{\dimexpr\textwidth-\fboxsep-\fboxrule\relax}{\copyrighttext}}};
\end{tikzpicture}%
}

\begin{document}

\title{Mixed Precision Solver Scalable to 16000 MPI Processes 
for Lattice Quantum Chromodynamics Simulations on the Oakforest-PACS System}

\author{\IEEEauthorblockN{
Taisuke Boku\IEEEauthorrefmark{1}\IEEEauthorrefmark{2},
Ken-Ichi Ishikawa\IEEEauthorrefmark{3}\IEEEauthorrefmark{4}
\thanks{\IEEEauthorrefmark{3}Email:ishikawa@theo.phys.sci.hiroshima-u.ac.jp}
\thanks{Submitted to LHAM'17 ``5th International Workshop on Legacy HPC Application Migration'' in
CANDAR'17 ``The Fifth International Symposium on Computing and Networking'' and to appear in the proceedings.},
Yoshinobu Kuramashi\IEEEauthorrefmark{5}\IEEEauthorrefmark{2} and
Lawrence Meadows\IEEEauthorrefmark{6}}
\IEEEauthorblockA{
\IEEEauthorrefmark{1}{Graduate School of Systems and Information Engineering, University of Tsukuba, Tsukuba, Ibaraki 305-8573, Japan}\\
\IEEEauthorrefmark{2}{Center for Computational Sciences, University of Tsukuba, Tsukuba, Ibaraki 305-8577, Japan}\\
\IEEEauthorrefmark{3}{Graduate School of Science, Hiroshima University, Higashi-Hiroshima, Hiroshima 739-8526, Japan}\\
\IEEEauthorrefmark{4}{Core of Research for the Energetic Universe, Hiroshima University, Higashi-Hiroshima, Hiroshima 739-8526, Japan}\\
\IEEEauthorrefmark{5}{Graduate School of Pure and Applied Sciences, University of Tsukuba, Tsukuba, Ibaraki 305-8571, Japan}\\
\IEEEauthorrefmark{6}{Intel Corporation, USA}
}}

\maketitle 
\pagestyle{plain}
\copyrightnotice

\begin{abstract}
Lattice Quantum Chromodynamics (Lattice QCD) is a quantum field theory 
on a finite discretized space-time box so as to numerically compute 
the dynamics of quarks and gluons to explore the nature of subatomic world. 
Solving the equation of motion of quarks (quark solver) is the most compute-intensive 
part of the lattice QCD simulations and is one of the legacy HPC applications.
We have developed a mixed-precision quark solver for 
a large Intel Xeon Phi (KNL) system named ``Oakforest-PACS'', 
employing the {\boldmath $O(a)$}-improved Wilson quarks as the discretized equation of motion.
The nested-BiCGSTab algorithm for the solver
was implemented and optimized using mixed-precision, 
communication-computation overlapping with MPI-offloading, SIMD vectorization, 
and thread stealing techniques.
The solver achieved {2.6 PFLOPS} in the single-precision part 
on a {\boldmath $400^3\times 800$} lattice using \num{16000} MPI processes on {8000} nodes on the system.
\end{abstract}

\section{Introduction}
\label{sec:Intro}

Proton and neutron, the constituents of atomic nuclei are not elementary particle. 
They are composed of three quarks, which are bound by the strong interaction through exchange of gluons.
The dynamics and interactions among quarks and gluons are described by 
Quantum Chromodynamics (QCD) which is a part of the Standard Model of the elementary particle physics.
The strong interaction shows a characteristic feature at a long distance scale larger than the radius of a proton: 
the quarks are ``confined'' inside proton/neutrons and never retrieved individually in any experiment.
This is completely nonperturbative phenomena so that it is not validated to make an analytic calculation of 
QCD based on the perturbative expansion in terms of the interaction strength. 
Lattice Quantum Chromodynamics (Lattice QCD)~\cite{Wilson:1974sk}, which is a quantum field theory defined
on a finite discretized space-time box, provides us a powerful tool to investigate the dynamics of quarks 
and gluons nonperturvatively in a numerical manner.

The most time consuming part of the Lattice QCD simulations is in solving the equation of motion of quarks.
The equation of motion of quarks is discretized to a large scale linear equation on a regular four-dimensional lattice.
Since the structure of the linear equation becomes a stencil type linear equation and 
the number of unknowns is typically $10^8$--$10^{10}$, the equation is usually solved by iterative algorithms 
such as Conjugate Gradient (CG) or Bi-Conjugate Gradient Stabilized (BiCGStab) algorithms.
The condition number of the coefficient matrix is inversely proportional to the quark mass.
As the mass of lightest quarks in nature is in $m_q\simeq 2$--$5$ [MeV] which is almost mass-less compared to 
the typical physical scale of QCD $\Lambda_{\mathrm{QCD}} \sim 1$ [GeV], which
implies a large condition number of $O(10^3)$ for the linear equation.
Thus tuning and optimizing the quark solver algorithm for upcoming High Performance Computing (HPC) architecture 
is the main challenge in lattice QCD simulations.

In this work we optimize the quark solver for the Wilson-Clover quarks~\cite{Sheikholeslami:1985ij} 
to the Oakforest-PACS (OFP) system~\cite{KNLSYSTEM:HOME}.  
The OFP system is made up of Intel Xeon Phi codenamed Knights Landing (KNL),
installed in the Kashiwa-no-Ha (Oakleaf) campus, the University of Tokyo, 
and operated by Joint Center for Advanced High Performance Computing (JCAHPC)~\cite{JCAHPC:HOME}
which is jointly established by the Center for Computational Sciences, University of Tsukuba (CCS)~\cite{CCS:HOME}
and the Information Technology Center, the University of Tokyo (ITC)~\cite{ITC:HOME}. 
To extract the best performance on the OFP system for the quark solver is the target of this work, 
and is a migration task of legacy HPC application. 
We focus on the weak-scaling performance of the quark solver using \num{16000} MPI processes on the OFP system. 
The lattice size of a four-dimensional lattice reaches to $400^3\times 800$, which is the largest to our current knowledge in the literature.
The vectorization using AVX-512 and OpenMP threading for many cores of Intel Xeon Phi (KNL) are the key issues of 
single process performance, which are implemented in the quark solver accordingly.
The communication overhead among the MPI processes is another issue to achieve the best performance
using such a large number of processes. 
We implemented the communication and computation overlapping algorithm combined with MPI-offloading
in the matrix-vector multiplication part of the quark solver.

This paper is organized as follows. 
We introduce the problem to be solved and the solver algorithm in section~\ref{sec:problem}.
The details of implementation and tuning are described in section~\ref{sec:implementation}.
The results of benchmarking are shown in~\ref{sec:results}. 
Section~\ref{sec:last} gives concluding remarks.

\section{Problem and Solver Algorithm}
\label{sec:problem}

\subsection{Problem to be Solved}

\begin{figure}[t]
  \centering
  \includegraphics[clip,scale=0.23,trim=0 15 0 15]{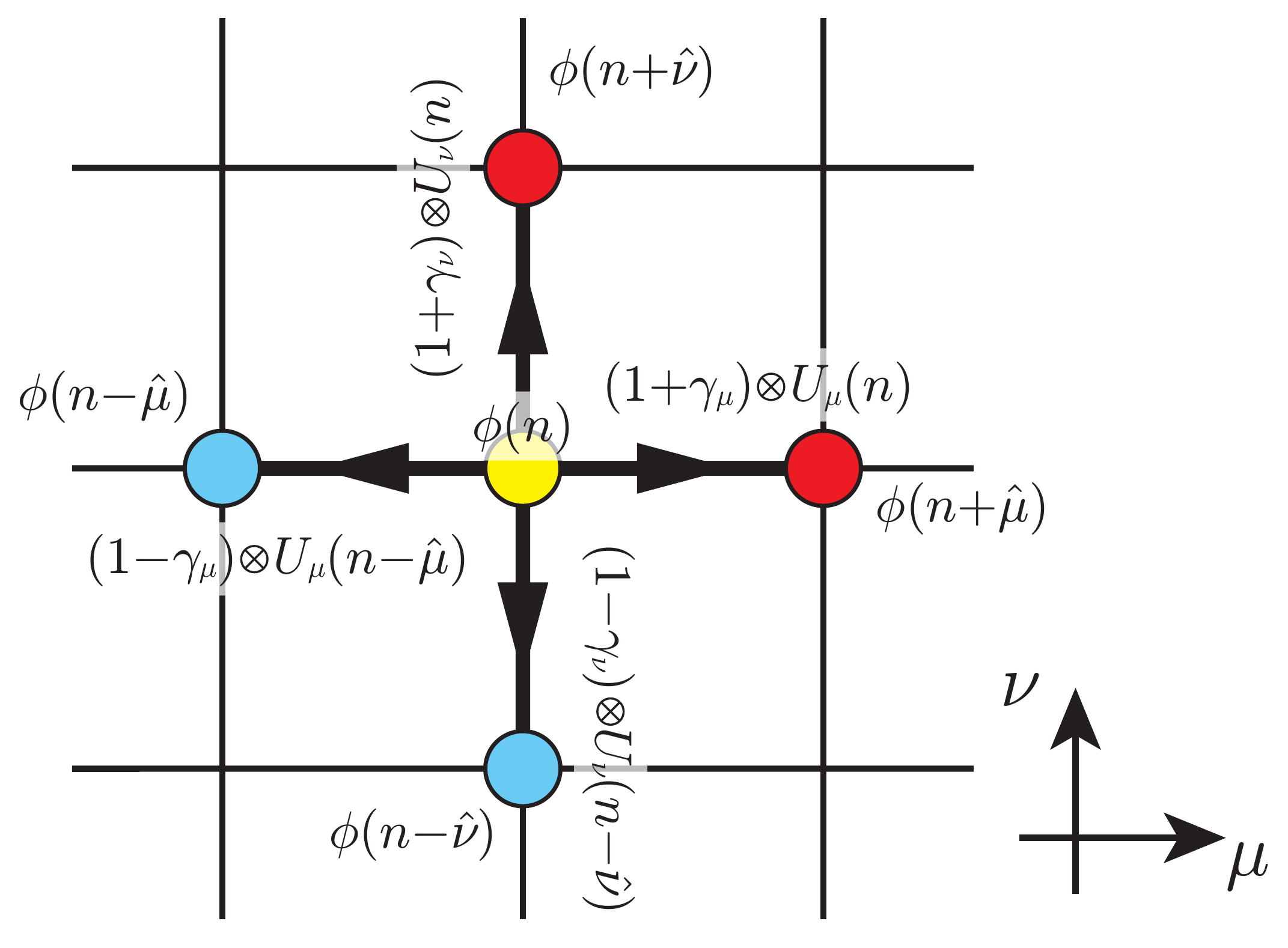}
  \vspace*{-0.5em}
  \caption{Hopping matrix stencil structure in the $\mu$--$\nu$ plane.}
  \label{fig:hopping}
\end{figure}

There are several types of the discretization forms for the equation of motion of quarks.
In this work we employ the $O(a)$-improved Wilson quark action~\cite{Sheikholeslami:1985ij}, 
which is simple and widely used from view points of the computational cost and the discretization error. 
The equation of motion is translated into the following linear equation;
\begin{linenomath}
\begin{align}
    D x &= b,
\label{eq:EOMQ}
\\
D &= \left(D^{i,j}_{\alpha,\beta}(n,m)\right),
 x = ( x^{i}_{\alpha}(n) ), 
 b = ( b^{i}_{\alpha}(n) ),
\end{align}
\end{linenomath}
where $x$ and $b$ are vectors containing the quark fields expressed with complex numbers,
$i$ is the color index running in $i=1,2,3$, $\alpha$ is the spinor index in $\alpha=1,2,3,4$,
and $n$ is the four-dimensional lattice site index $n=(n_x,n_y,n_z,n_t)$. 
$D$ is called the Wilson-Clover matrix.  It is a sparse matrix defined by
\begin{linenomath}
\begin{align}
D^{i,j}_{\alpha,\beta}(n,m) & =
F^{i,j}_{\alpha,\beta}(n)\delta_{n,m}
-\kappa 
\mathrm{H}^{i,j}_{\alpha,\beta}(n,m),
\label{eq:WilsonCloverOP}
\\
\mathrm{H}^{i,j}_{\alpha,\beta}(n,m) & \equiv
\!\!\!\!
\sum_{\mu=x,y,z,t}
\!\!\!\!
\left[
 (1-\gamma_{\mu})_{\alpha,\beta} U_{\mu}(n)^{i,j}
\delta_{n+\hat{\mu},m}
\right.
\notag\\
&\quad\left.
+(1+\gamma_{\mu})_{\alpha,\beta} (U_{\mu}(m)^{j,i})^{*}
\delta_{n-\hat{\mu},m}
\right],
\label{eq:Hopping}
\end{align}
\end{linenomath}
where $\delta_{n,m}$ is the four-dimensional Kronecker delta,
$\hat{\mu}$ is a unit vector indicating the $\mu$-direction, and
$\gamma_\mu$'s are constant $4\times 4$ Hermitian matrices called 
Dirac's gamma matrices where $\gamma_t$ is defined in diagonal.
$U_{\mu}(n)^{i,j}$ is the gauge (gluon) field forming the matrix-valued four-vector 
field as $U_{\mu}(n)=(U_{\mu}(n)^{i,j})$. 
$F_{\alpha,\beta}^{i,j}(n)$ is a matrix field made of $U_{\mu}(n)$.
$\kappa$ and $\mathrm{H}^{i,j}_{\alpha,\beta}(n,m)$ are called as 
the hopping parameter and hopping matrix, respectively. $\kappa$ is inversely related to the quark mass.
$F(n)=(F_{\alpha,\beta}^{i,j}(n))$ is a $12\times 12$ Hermitian matrix called \textit{Clover} term, 
and $U_{\mu}(n)$ is a $3\times 3$ special unitary matrix.
The quark fields are located on the lattice sites and
the gauge fields reside on the links connecting nearest neighboring sites
(Figure~\ref{fig:hopping}).  
Periodic boundary condition is imposed in all directions.

Equation~(\ref{eq:EOMQ}) is solved many times during updating the gauge field $U_{\mu}(n)$ to 
produce the quantum mechanical expectation value of the observables made of quarks.
Typically $10^{6}$--$10^{8}$ solutions are needed in the lattice QCD simulations.

\subsection{Structure of the algorithm}

Equation~(\ref{eq:EOMQ}) is preconditioned with the even-odd (or red/black) site ordering
after the diagonal block preconditioning for the local clover term $F(n)$. 
The equation is then transformed into the following form.
\begin{linenomath}
\begin{align}
    \hat{D}_{ee} x_{e} &= \hat{b}_e,
\label{eq:EOpreconditionedEQ}
\\
\hat{b}_e&= F_{ee}^{-1} b_{e} + \kappa \tilde{\mathrm{H}}_{eo}F_{oo}^{-1}b_{o},\\
      x_o&= F_{oo}^{-1} b_{o} + \kappa \tilde{\mathrm{H}}_{oe}x_{e},\\
\hat{D}_{ee} &\equiv 1_{ee} - \kappa^2 \tilde{\mathrm{H}}_{eo}\tilde{\mathrm{H}}_{eo},\\
\tilde{\mathrm{H}}_{eo}&\equiv F_{ee}^{-1}\mathrm{H}_{eo}, \quad 
\tilde{\mathrm{H}}_{oe} \equiv F_{oo}^{-1}\mathrm{H}_{oe},
\end{align}
\end{linenomath}
where equation~(\ref{eq:EOpreconditionedEQ}) is to be solved with the nested-BiCGStab 
algorithm~\cite{Vogel2007226,Tadano2008} combined with mixed-precision technique.
The vectors with subscript $e$ ($o$) contains the field elements on 
even (odd) sites labeled by $n_x+n_y+n_z+n_t \mod 2 = 0$ ($n_x+n_y+n_z+n_t \mod 2 = 1$) only.
$\mathrm{H}_{eo}$ is the hopping matrix connecting sites from odd sites to even sites and vice versa for $\mathrm{H}_{oe}$.
The clover term $F_{ee}$ ($F_{oo}$) is diagonal in terms of site index and can be inverted easily.

As the matrix $\tilde{\mathrm{H}}_{eo}$ ($\tilde{\mathrm{H}}_{oe}$) is
the key  building block in the solver for the $O(a)$-improved Wilson quark, 
we call the matrix $\tilde{\mathrm{H}}_{eo}$ (and $\tilde{\mathrm{H}}_{oe}$) as 
\textbf{MULT} in this paper. 
The code for the matrix-vector multiplication with \textbf{MULT} is one of the optimization target.

In order to utilize the cache memory and reduce the memory bandwidth requirement 
the single-precision BiCGStab algorithm is used as the inner-solver of 
the flexible-BiCGStab algorithm. The algorithm is summarized in 
Algs.~\ref{alg:Outerloop} and \ref{alg:InnerloopSP}.
Entire timing is measured as indicated in Alg.~\ref{alg:Outerloop} 
and the detailed performance is only monitored for the inner-solver as shown 
in Alg.~\ref{alg:InnerloopSP}.

We employ the CCS QCD Solver Benchmark program publicly available from~\cite{CCSQCD},
which implements the BiCGStab algorithm, as the base code of our benchmark program.
The original code of~\cite{CCSQCD} is solely written with Fortran90 in double precision.
We implemented and added the nested-BiCGStab algorithm (Algs.~\ref{alg:Outerloop} and \ref{alg:InnerloopSP}) 
to the original code. 
This approach (adding only the single precision solver) minimizes code modification cost.
The single precision part is written with C/C++ language 
to use the SIMD intrinsic functions for AVX-512 using the Intel Compiler suite.

\begin{figure}[t]
  \removelatexerror
  \begin{algorithm}[H]
  \caption{\algcapsize Solve $D x = b$ for $x$ using the nested-BiCGStab algorithm. 
  Symbols with $(\mathrm{SP})$ are the data arrays in single precision.}
  \label{alg:Outerloop}
  \begin{algorithmic}[1]
        \State{Start timer.}
        \State{Convert $[U_{\mu}(n),F(n)]$ to $[U^{\SP}_{\mu}(n),F^{\SP}(n)]$.}
        \State{$\hat{b}_e = F_{ee}^{-1} b_{e} + \kappa \tilde{\mathrm{H}}_{eo}F_{oo}^{-1}b_{o}$.}
        \State{$x_e=0; r_e=\tilde{r}_e=p_e=\hat{b}_e$.}
        \State{$\mathrm{src}=|\hat{b}_e|; \rho_0 = \mathrm{src}^2$}
        \For{$\mathrm{iter}=0$ to $\mathrm{iter}_{\mathrm{MAX}}$}
          \State{$v_e = M_{ee} p_e$ (call Alg.~\ref{alg:InnerloopSP})}
          \State{$q_e = \hat{D}_{ee} v_e $}
          \State{$\alpha=\langle \tilde{r}_e| q_e \rangle$}
          \State{$x_e = x_e + \alpha v_e$; $r_e = r_e - \alpha q_e$}
          \State{\textbf{if} {$|r_e|/\mathrm{src} < \mathrm{tol}$} \textbf{exit}}
          \State{$v_e = M_{ee} r_e$ (call Alg.~\ref{alg:InnerloopSP})}
          \State{$t_e = \hat{D}_{ee} v_e$}
          \State{$\omega=\langle t_e |r_e \rangle/|t_e|^2$}
          \State{$x_e = x_e + \omega v_e$; $r_e = r_e - \omega t_e$}
          \State{\textbf{if} {$|r_e|/\mathrm{src} < \mathrm{tol}$} \textbf{exit}}
          \State{$\rho_1 = \langle\tilde{r}_e|r_e \rangle$;
                 $\beta=(\rho_1/\rho_0)/(\alpha/\omega)$;
                 $\rho_0 = \rho_1$}
          \State{$p_e = p_e -\omega q_e; p_e = r_e + \beta p_e$}
        \EndFor
        \State{$x_o = F_{oo}^{-1} b_{o} + \kappa \tilde{\mathrm{H}}_{oe}x_{e}$.}
        \State{Stop timer.}
  \end{algorithmic}
  \end{algorithm}
\vspace*{-1em}
  \begin{algorithm}[H]
  \caption{\algcapsize Inner solver for $v_e = M_{ee} p_e$.}
  \label{alg:InnerloopSP}
  \begin{algorithmic}[1]
        \State $\mathrm{norm}=|p_e|$
        \State $b^{\SP}_e=p_e/|p_e|$
        \State Start performance monitoring.
        \State $x^{\SP}_e=0; r^{\SP}_e=\tilde{r}^{\SP}_e=p^{\SP}_e=b^{\SP}_e$.
        \State $\rho^{\SP}_0 = 1$
        \For{$\mathrm{iter}=0$ to $\mathrm{iter}_{\mathrm{MAX}}$}
        \State $q^{\SP}_e = \hat{D}^{\SP}_{ee} p^{\SP}_e $
        \State $\alpha^{\SP}=\langle \tilde{r}^{\SP}_e| q^{\SP}_e \rangle$
        \State $x^{\SP}_e = x^{\SP}_e + \alpha^{\SP} p^{\SP}_e$; $r^{\SP}_e = r^{\SP}_e - \alpha^{\SP} q^{\SP}_e$
        \State \textbf{if} {$|r^{\SP}_e| < \mathrm{tol}$} \textbf{exit}
        \State $t^{\SP}_e = \hat{D}^{\SP}_{ee} r^{\SP}_e$
        \State $\omega^{\SP}=\langle t^{\SP}_e |r^{\SP}_e \rangle/|t^{\SP}_e|^2$
        \State $x^{\SP}_e = x^{\SP}_e + \omega^{\SP} r^{\SP}_e$; $r^{\SP}_e = r^{\SP}_e - \omega^{\SP} t^{\SP}_e$
        \State \textbf{if} {$|r^{\SP}_e| < \mathrm{tol}$} \textbf{exit}
        \State{$\rho^{\SP}_1 = \langle\tilde{r}^{\SP}_e|r^{\SP}_e \rangle$;
               $\beta^{\SP}=(\rho^{\SP}_1/\rho^{\SP}_0)/(\alpha^{\SP}/\omega^{\SP})$;
               $\rho^{\SP}_0 = \rho^{\SP}_1$}
        \State $p^{\SP}_e = p^{\SP}_e -\omega^{\SP} q^{\SP}_e; p^{\SP}_e = r^{\SP}_e + \beta^{\SP}  p^{\SP}_e$
        \EndFor
        \State Stop performance monitoring.
        \State $v_e=\mathrm{norm}\times x^{\SP}_e$
  \end{algorithmic}
  \end{algorithm}
\end{figure}

\subsection{Related Works}

Here we summarize the current status of the performance of the solver for 
the Wilson-Clover quarks in the following. 

\subsubsection{NVIDIA GPU systems}
A lattice QCD library on GPUs named \textbf{QUDA} is available in~\cite{QUDA:HOME} and 
the details of the implementation and the performance are described in~\cite{Clark:2009wm,Babich:2011np}.
The performance for the Wilson-Clover matrix multiplication and the solver has been investigated 
on a cluster system equipped with NVIDIA Tesla M2050 GPUs~\cite{Babich:2011np}, where
a good strong-scaling for the GCR algorithm solver with a domain-decomposition (DD) preconditioner 
in mixed (half-single) precision was observed up to 256 GPUs 
on a $32^3\times 256$ lattice, reaching over 16 TFLOPS on the system. 
An update on the Wilson-Clover matrix kernel performance has been presented by M.~Wagner 
in the lattice conference (Lattice 2016)~\cite{NVIDIA:Wagner:LAT2016}, 
where the kernel performance on a $32^4$ lattice reaches 0.8--1 TFLOPS (1.5--2.0 TFLOPS) 
in single (half) precision using single P100 card.

\subsubsection{Intel Xeon Phi Coprocessors (Knights Corner: KNC)}
The hopping matrix kernel performance on Intel Xeon Phi Coprocessors (Knights Corner: KNC)
has been investigated in~\cite{JOO:ISC2013}.
Using single precision arithmetic, they achieved 320 GFLOPS for the kernel 
and 237 GFLOPS for the CG solver on a $32\times 40\times 24\times 96$ lattice 
using single coprocessor card, and observed a good strong-scaling property 
on $32^3\times 256$ and $48^3\times 256$ lattices up to 32 cards.
In~\cite{Heybrock:2014iga}, a DD preconditioner has been applied to enhance 
the strong-scaling and the convergence performance.
They observed a better strong-scaling behavior up to 1024 cards. 
The performance observed in~\cite{Heybrock:2014iga} was 400--500 GFLOPS on a card.
A similar study has been performed in~\cite{Boku:2016dmw} and it was
observed that the lattice size on a card must be larger than $24^3\times 32$ 
to obtain a good weak-scaling property.

\subsubsection{Intel Xeon Phi (Knights Landing: KNL) systems}
B.~Jo{\'o} \textit{et al.}~\cite{Joo2016KNL} have reported the performance with the Wilson-Clover quarks on KNL
with the code developed based on the publicly available \textbf{QPhiX} code~\cite{QPhiX:HOME}.
The Wilson and Wilson-Clover kernels are investigated on both 
a Xeon Haswell system and a KNL system, and the weak-scaling performance is measured
up to 16 sockets with a $32^4$ lattice per MPI-RANK. 
The CG solver with the Wilson-Clover kernel performs 3.5--4.0 TFLOPS on 16 sockets 
in the weak-scaling benchmark.

The quark solver performance achieved in the past few years is reviewed 
by P.~Boyle in~\cite{Boyle:2017wul}.

\section{Implementation details}
\label{sec:implementation}

We explain the details of the optimization on \textbf{MULT} ($\tilde{\mathrm{H}}_{eo}^{\SP}$) 
in this section. 

\subsubsection{SIMD vectorization}
\begin{figure}[t]
    \centering
    \includegraphics[clip,scale=0.25]{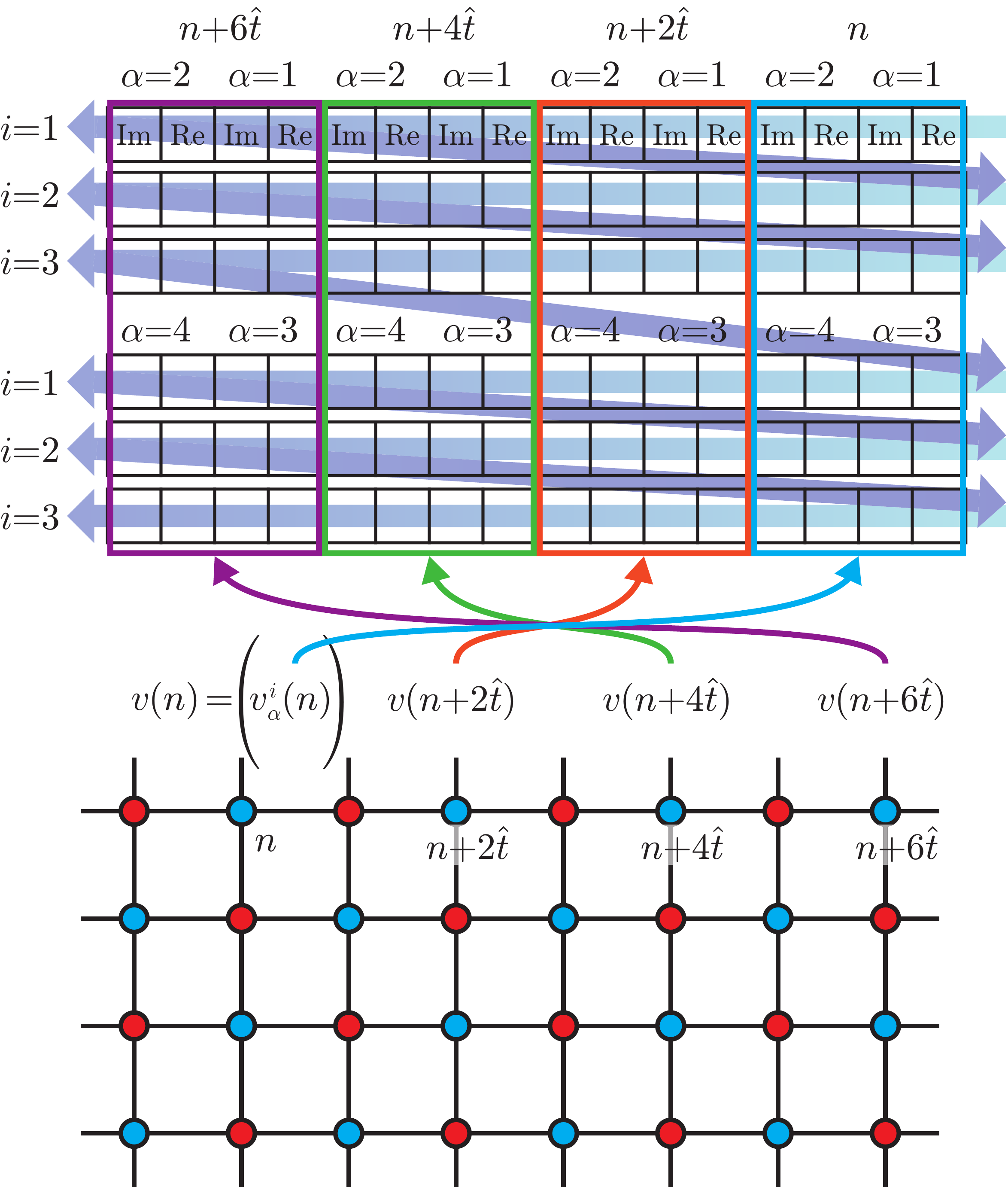}
    \caption{Memory layout of the quark field in single precision. 
   Even-odd ordering is employed for the lattice indexing. 
   The horizontal direction on the lattice figure corresponds to the temporal direction and 
   the vertical one to the spatial direction.
   $\mathrm{Re}$ and $\mathrm{Im}$ indicate a real and an imaginary part 
   of a field component $v^i_\alpha(n)$ respectively.}
    \label{fig:SIMDlayoutquark}
\end{figure}

We employ the memory layout  shown in Figure~\ref{fig:SIMDlayoutquark}
for quark fields to fit them in the SIMD vectorization 
in single precision.  
The multiplication of $(1\pm\gamma_\mu)$ (linear operation on the spinor index $\alpha$) 
is thus done within a lane of SIMD vectors 
(there are 4 lanes in AVX-512 and single lane corresponds to a SSE vector).

The memory layout for the gauge field $U_{\mu}^{\SP}(n)$ is similar to 
that of the spinor field after dropping the third column of the $3\times 3$ 
matrix of $U_{\mu}(n)$ using the property of special unitary matrix. 
The indexing on the color is rephrased as 
$ U_{\mu}(n)^{i,\alpha}(n) \to v^{i}_{\alpha}(n)$ with $\alpha=1$ and $2$ only.
Thus the gauge field can be treated as the two-component spinor.
The missing third column is recomputed as $U=(\vec{a},\vec{b},(\vec{a}\times \vec{b})^{*})$
where $\vec{a}$ and $\vec{b}$ are complex-valued column vectors in three-dimension.
This is called \textbf{SU(3)-reconstruction} technique, which reduces the memory footprint and 
improves the cache usability.

The clover term $F(n)$ can be reduced to two $6\times 6$ Hermitian matrices.
The $36\times 2$ reals of two $6\times 6$ Hermitian matrices from 
four time-slices are also packed in a memory layout similar to the spinor field.
Thus all arithmetic operations on the fields
needed in the solver Alg.~\ref{alg:InnerloopSP} are written with 
the AVX-512 SIMD intrinsic functions.

The number of floating-point operations (FLOP) per lattice site is summarized in Table~\ref{tab:flopcounts},
where the FLOP counts/site for \textbf{MULT} involves the overhead of \textbf{SU(3)-reconstruction} technique.
The performance of the single precision solver is estimated from 
the FLOP counts and the timing measured in the algorithm.
\begin{table}[t]
    \centering
    \begin{threeparttable}
    \centering
    \caption{Flop count per site in each function}
    \label{tab:flopcounts}
    \begin{tabular}{lr}\hline
Function\tnote{1} & FLOP counts/site  \tabularnewline\hline\hline
$w_e = \tilde{\mathrm{H}}_{eo} v_o$ (\textbf{MULT})  & {2232} \\
$q_e = \hat{D}_{ee} p_e$
& {4512} \\
$w_e = w_e + \alpha v_e$  or $w_e = v_e + \alpha w_e$ &   96 \\
                    $\delta = \langle v_e|w_e\rangle$ &   96 \\
                             $\zeta = |v_e|^2$        &   48 \\
\hline
   \end{tabular}
   \begin{tablenotes}
   \item[1]$\alpha$ and $\delta$ are complex numbers.
   \end{tablenotes}
   \end{threeparttable}
\end{table}

\subsubsection{Loop optimization for cache and threading}
The total lattice size is parametrized by $N_{X}\times N_{Y} \times N_{Z} \times N_{T}$
and the MPI-RANK size with the four-dimensional partitioning by 
$P_X\times P_Y\times P_{Z}\times P_{T}$.
The local lattice size per MPI-RANK is $N_{LX}\times N_{LY} \times N_{LZ} \times N_{LT}$ 
which satisfies $N_{Lj}\times P_j=N_{j}, (j=X,Y,Z,T)$.
We use equal local extent in the spatial directions as $N_{LX}=N_{LY}=N_{LZ}=N_{LS}$.

The four-dimensional loop visiting the lattice sites is tiled to extract 
the best performance from many KNL cores.  
The site indexing is $(n_x,n_y,n_z,nb_t)$ in the four time-slices major ordering followed by 
the even/odd-site ordering, where $nb_t$ is the block index of the bunch of four time-slices.
We decompose the four-dimensional loop into several small loops of 
$2\times 2\times 4\times 8$ ($x \times y\times z\times t$) by loop tiling,
and the data size on the small lattice is estimated to be 336 KiB
which fits in the 512KiB size of L2 cache of a physical core
as the two cores share the 1MB L2 cache in the tile of KNL architecture.
The prefetching intrinsics are inserted in the kernel loop body accordingly.

Two SMT threads are assigned to the temporal loop in the tile while
others are assigned to the block index using OpenMP. 
When the lattice size in a process is not a multiple of the tile 
size of $2\times 2\times 4\times 8$, there will be a thread imbalance 
to process the reminder sites.
The thread parallel performance on the block index is optimized using
a work stealing scheduling technique to minimize the thread imbalance.

\begin{figure}[t]
    \centering
    \includegraphics[clip,scale=0.40,trim=0 0 0 0]{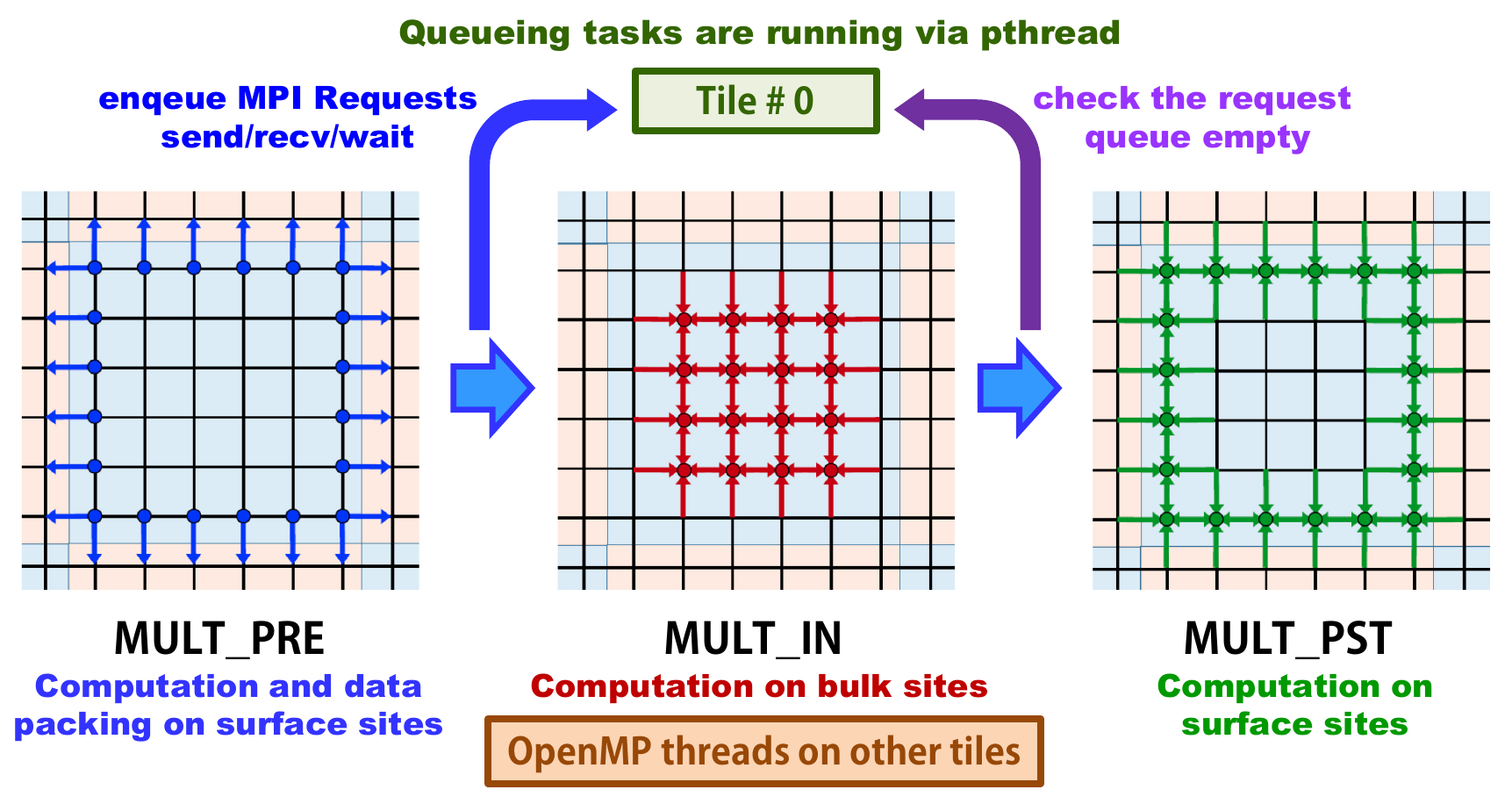}
    \caption{\textbf{MULT} operation in three steps.}
    \label{fig:MULTSPLITKNL}
\end{figure}

\newcommand{\figscale}{0.40}
{\setlength{\tabcolsep}{0pt}
\begin{figure*}[t]
    \centering
    \begin{tabular}{ccc}
    \includegraphics[clip,scale=\figscale]{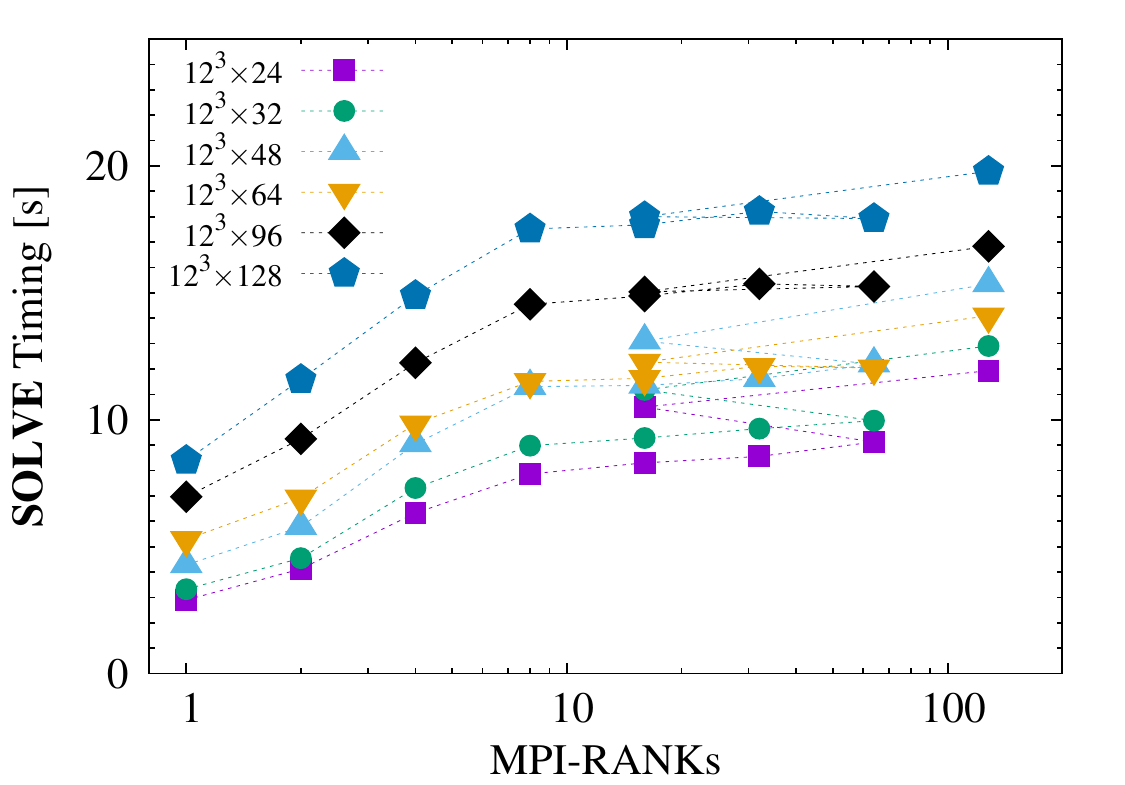} &
    \includegraphics[clip,scale=\figscale]{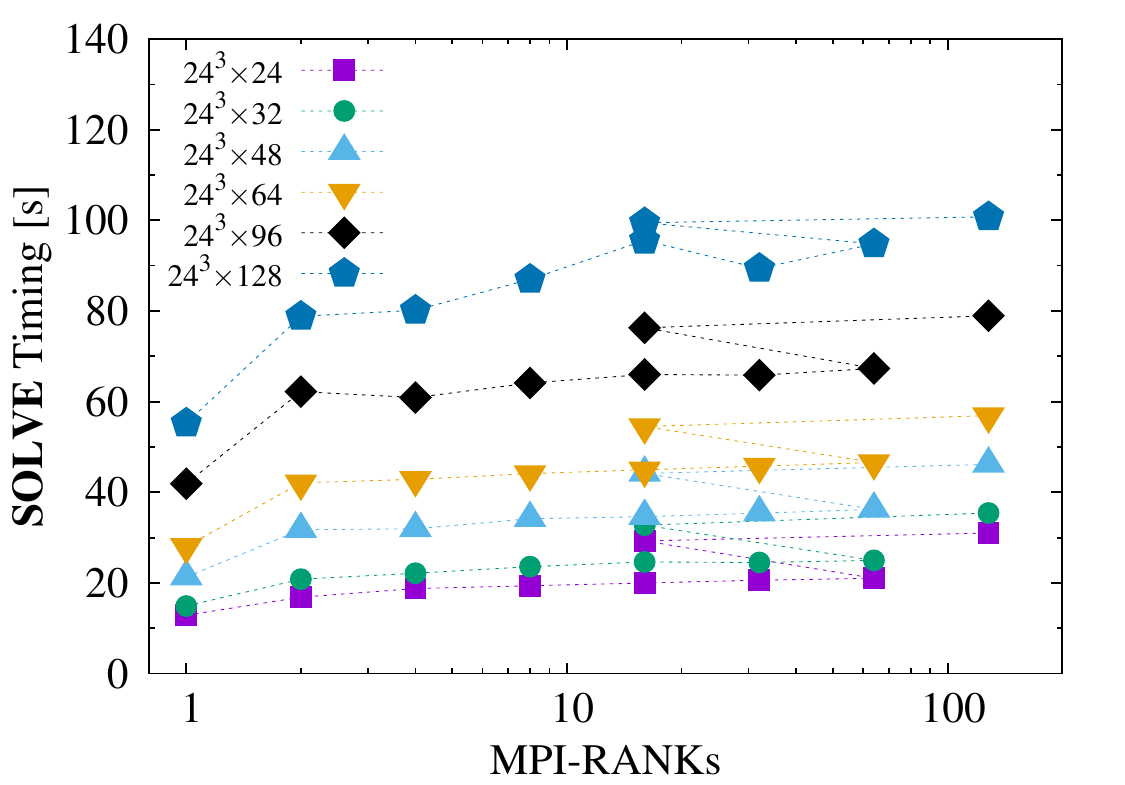} &
    \includegraphics[clip,scale=\figscale]{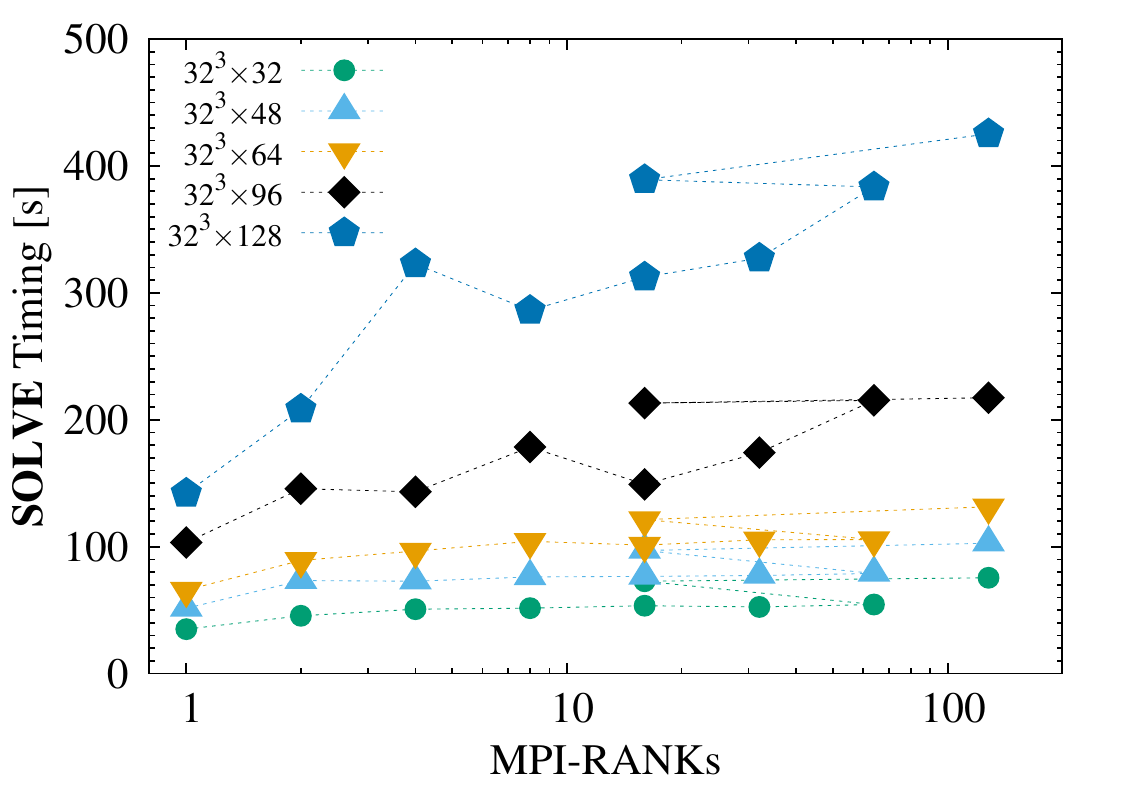}\\
    \includegraphics[clip,scale=\figscale]{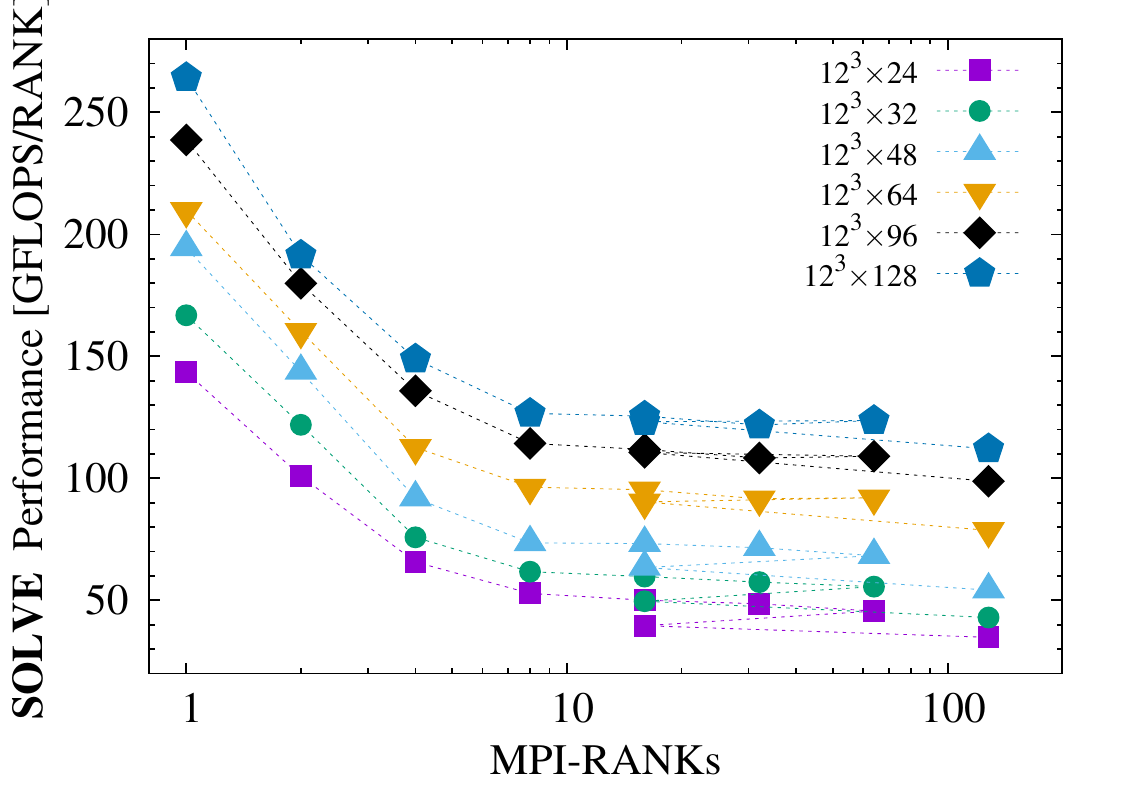} &
    \includegraphics[clip,scale=\figscale]{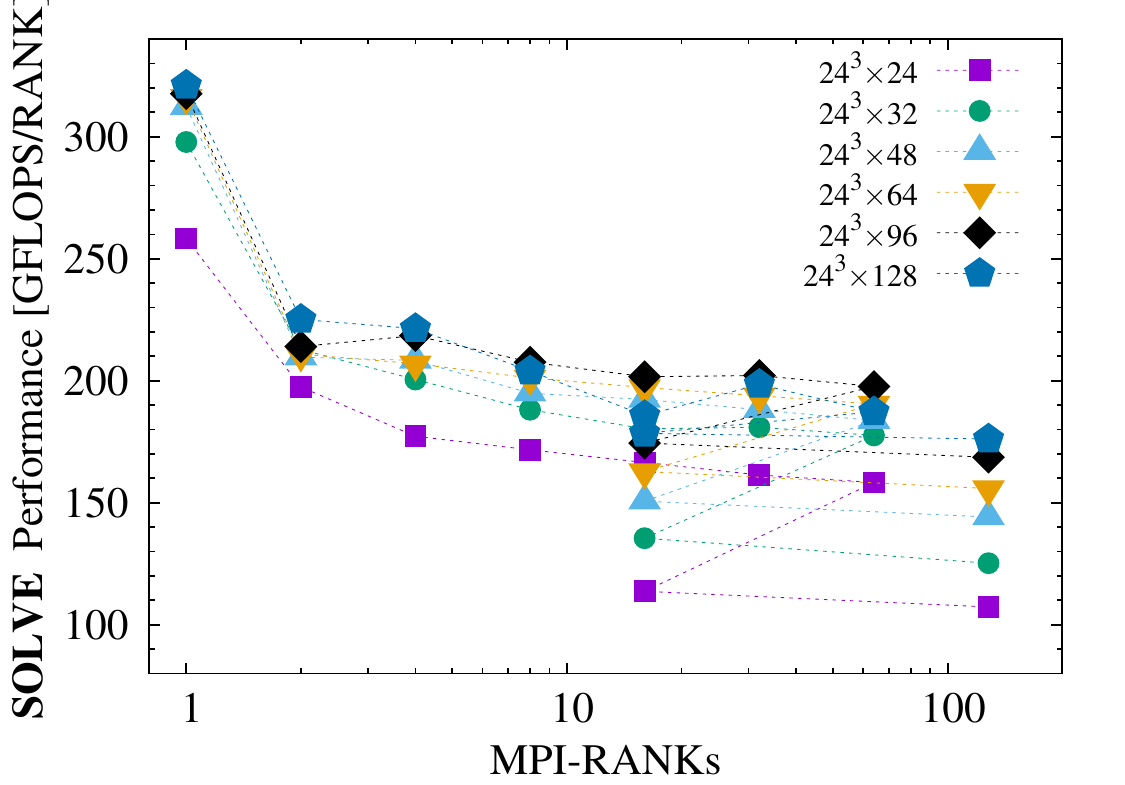} &
    \includegraphics[clip,scale=\figscale]{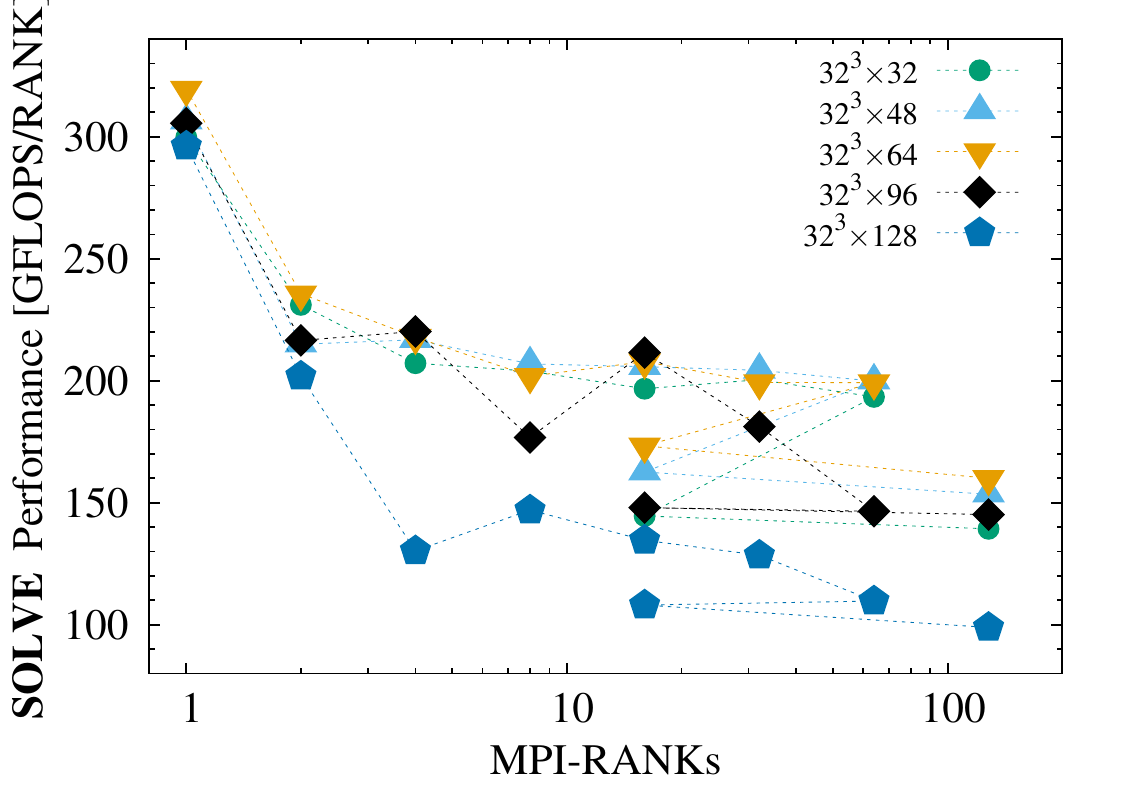}
    \end{tabular}
    \caption{Weak-scaling test results: \textbf{SOLVE} timing (upper figs) and performance (lower figs). 
             $N_{LS}=12, 24$, and 32 are shown.}
    \label{fig:SOLVE_TIME_SMALL_WS}
\end{figure*}}

\subsubsection{Communication-computation overlapping and MPI offloading}
We use {8000} nodes of the OFP system for the lattice simulation on a $400^3\times 800$ lattice.
We partition the lattice into $20^3\times 2 =$ \num{16000} MPI processes 
so that each MPI process treats a $20^3\times 400$ local lattice.
We put two MPI processes on a node. 
Reducing the overhead of the MPI communications is important for such a huge number 
of MPI processes.
In~\cite{Boku:2016dmw}, we implemented a communication-computation overlapping 
technique on the KNC system where the host CPU is assigned 
for the MPI-offloading while the KNC card computes the kernel. 
As the OFP system we examined is not host CPU+accelerator card architecture, 
we need another approach to offload MPI functions.

To hide the communication behind the computation we split 
\textbf{MULT} into three steps;
\textbf{MULT\_PRE},
\textbf{MULT\_IN}, and
\textbf{MULT\_PST}; as shown in Figure~\ref{fig:MULTSPLITKNL}.
The local lattice sites are classified into bulk sites and surface sites, where
the stencil computation for the former is done 
in \textbf{MULT\_IN} without 
waiting for the data from neighboring MPI-RANKs.
Data-packing after multiplications of $(1+\gamma_\mu)U_{\mu}(n)^{\dag}$ and $(1-\gamma_\mu)$ 
is done in \textbf{MULT\_PRE} and the reminder of the stencil computations after unpacking data received is done in \textbf{MULT\_PST}.
During the computation in \textbf{MULT\_IN} the surface data can be transmitted to other ranks provided 
that MPI functions are enabled to be concurrent to the computation.

In order to realize the concurrency in processing MPI functions and computing the kernel \textbf{MULT\_IN}
we assign the cores located in the tile \#0 of the KNL chip for MPI functions using \textit{pthread} mechanism.
The processor set on the tile \#0 are (0,1,68,69,136,137,204,205), 
corresponding two physical cores with 4 SMT threads enabled, in the case of 
Intel Xeon Phi 7250 (KNL:68 cores). These processors are excluded from 
the process pinning of the Intel MPI so that the compute kernels run only on 
the cores excluding the cores on the tile \#0.
To offload MPI functions to the processors on the tile \#0, 
we call \texttt{pthread\_create} on the processor set 
to create a thread monitoring and processing a queue to which
MPI functions are submitted from the master thread of 
the computing kernels.
With this mechanism we safely execute the MPI functions and the computation concurrently.

{
\setlength{\tabcolsep}{0pt}
\begin{figure*}[t]
    \centering
    \begin{tabular}{ccc}
    \includegraphics[clip,scale=\figscale]{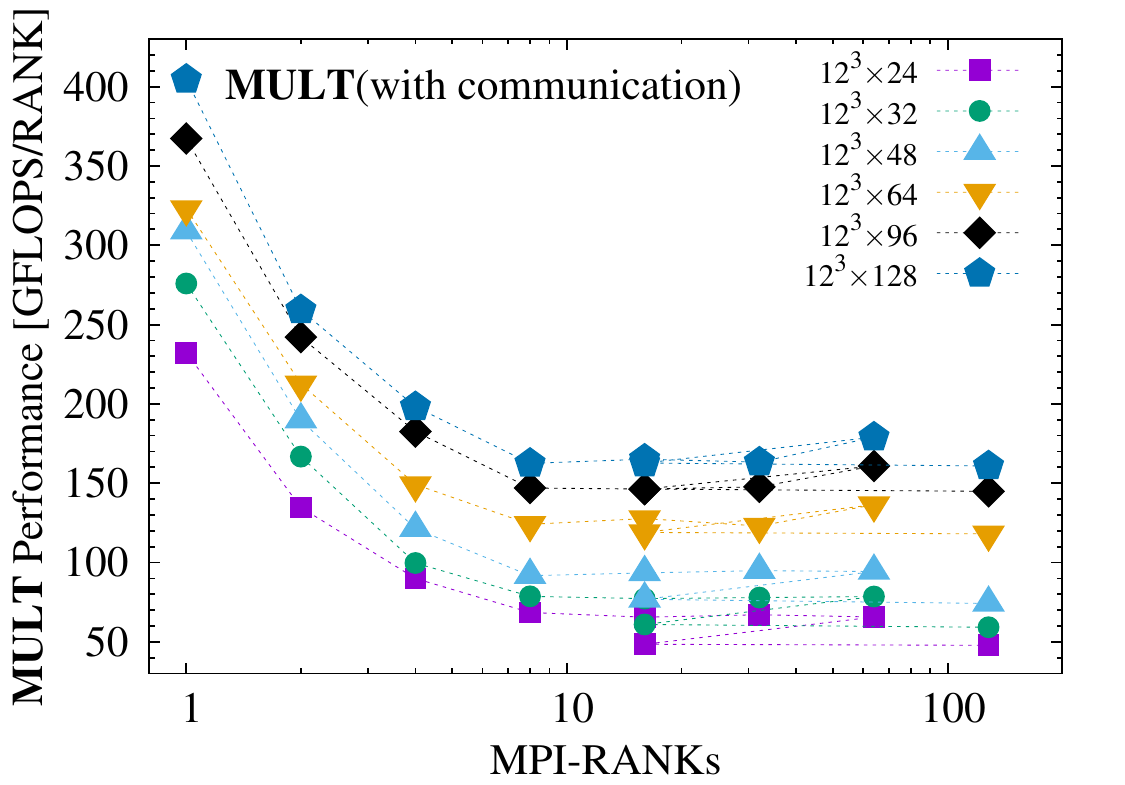} &
    \includegraphics[clip,scale=\figscale]{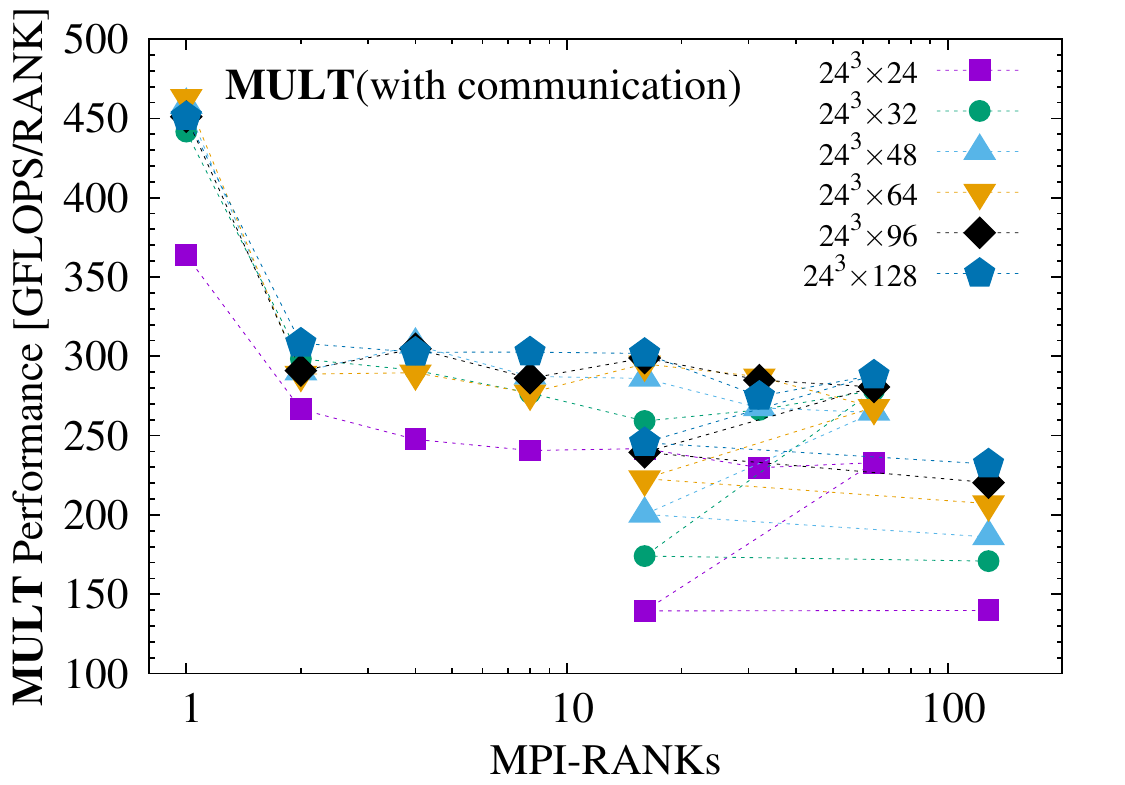} &
    \includegraphics[clip,scale=\figscale]{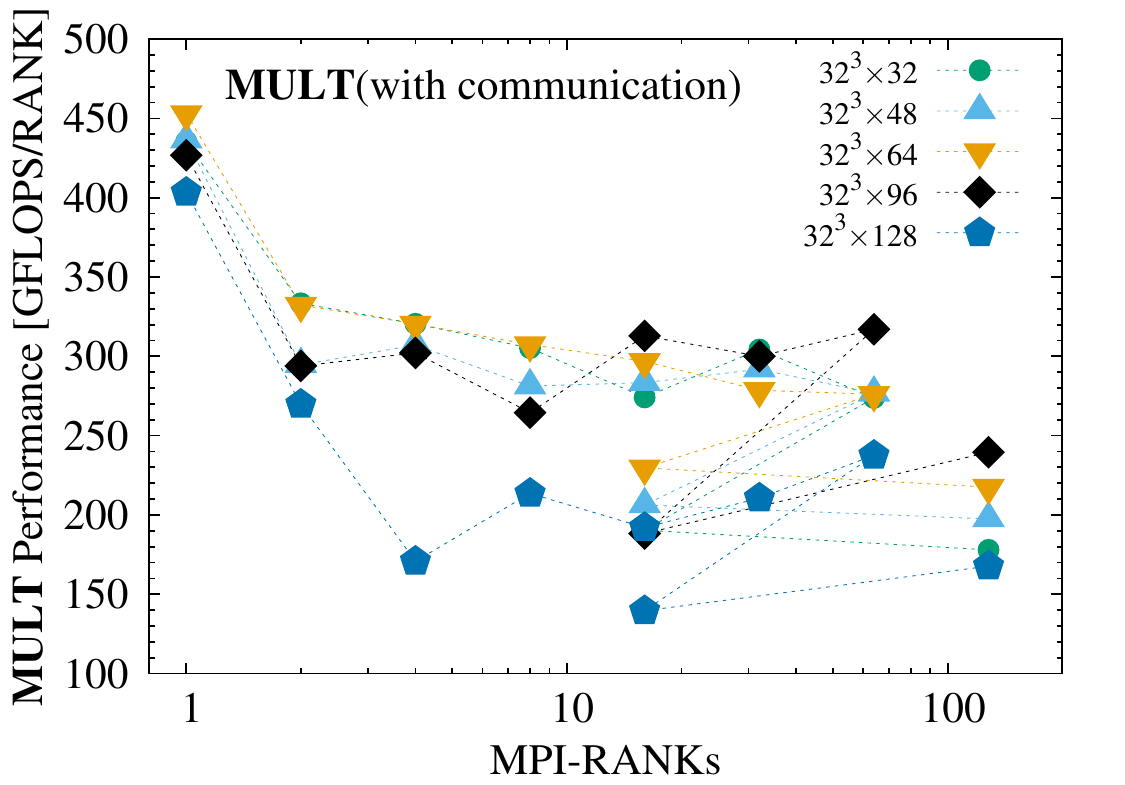}\\
    \includegraphics[clip,scale=\figscale]{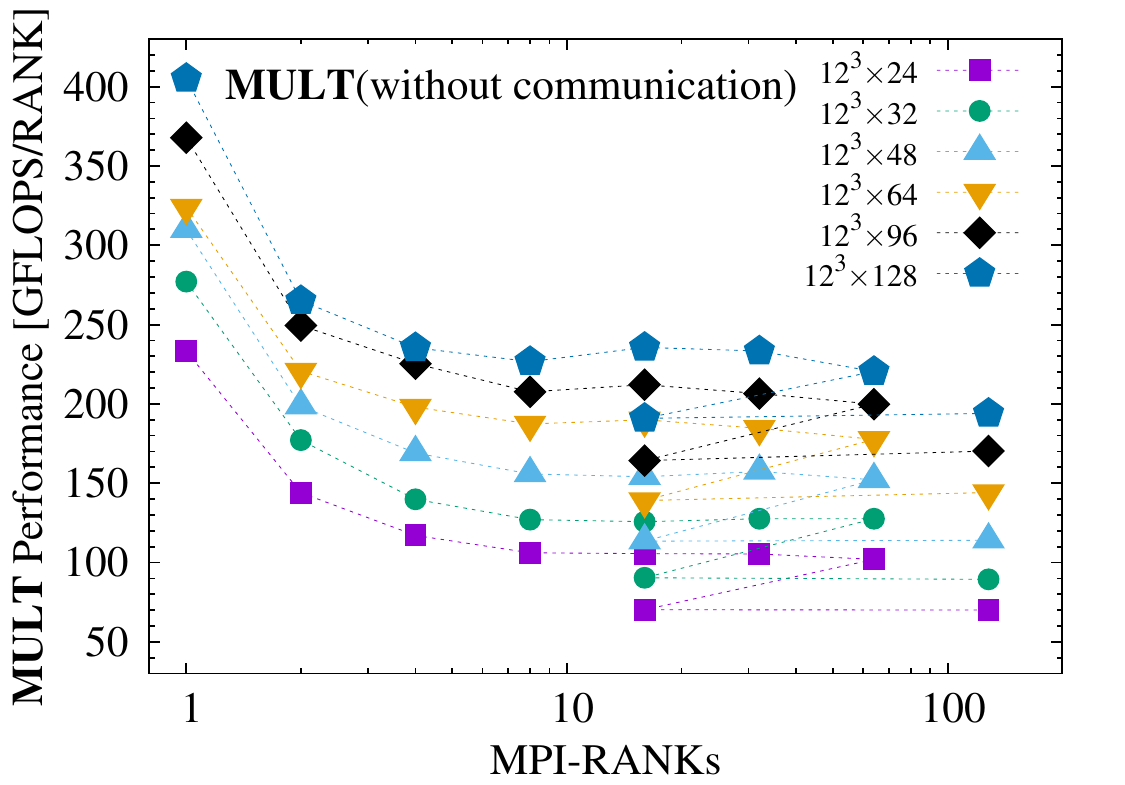} &
    \includegraphics[clip,scale=\figscale]{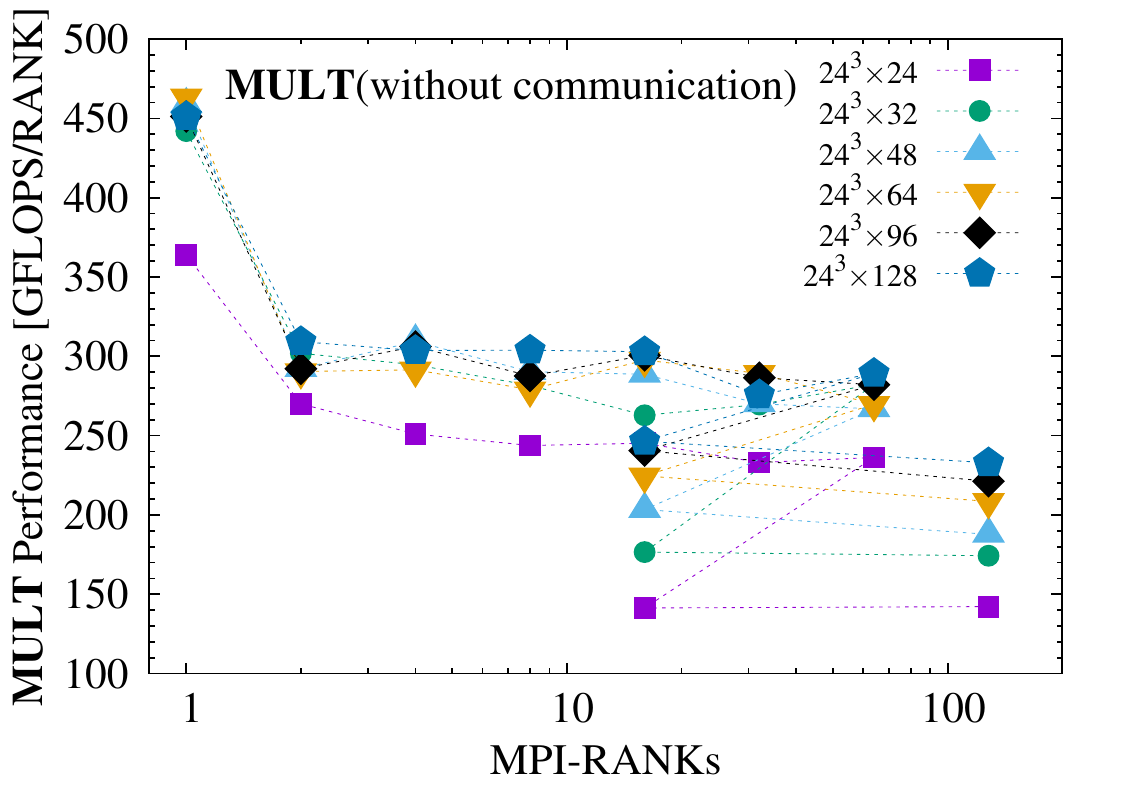} &
    \includegraphics[clip,scale=\figscale]{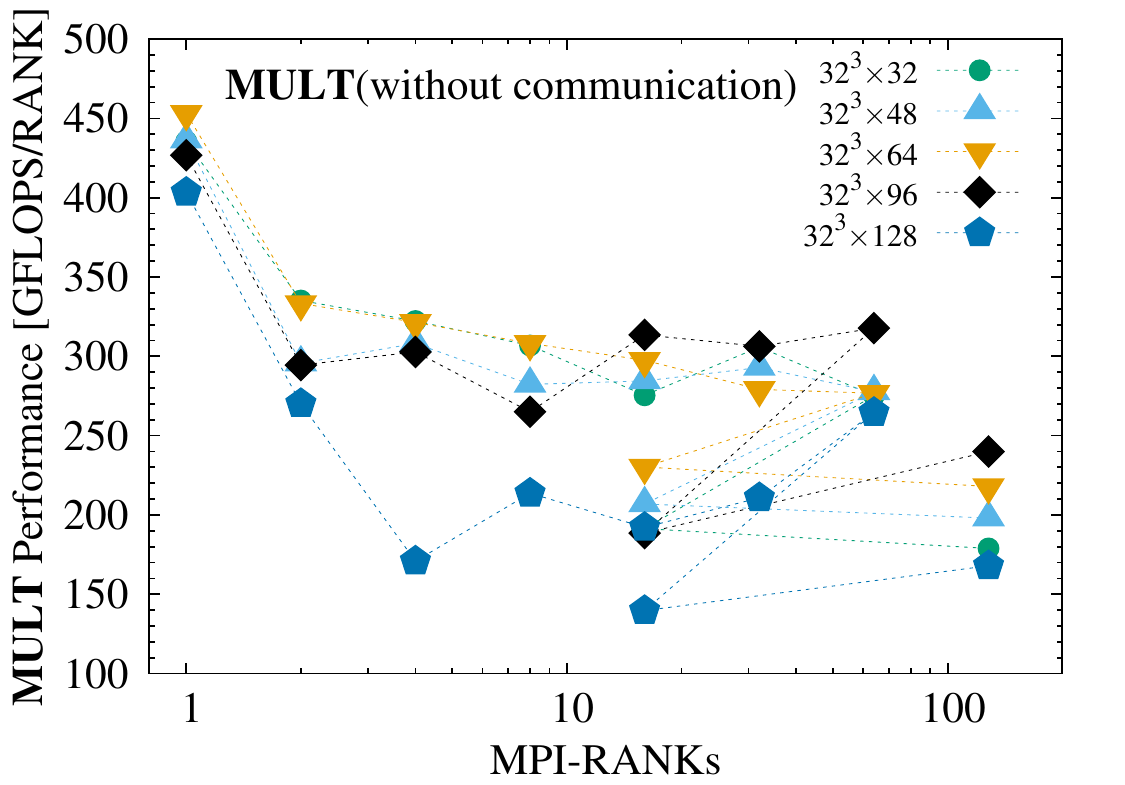}
    \end{tabular}
    \caption{Weak-scaling test results: \textbf{MULT} performance with comm. (upper figs) and without comm. (lower figs).}
    \label{fig:MULT_PERF_SMALL_WS}
\end{figure*}
}

\section{Benchmarking Results}
\label{sec:results}

\subsection{Oakforest-PACS System Overview}
\label{sec:OFPsystem}

The Oakforest-PACS System~\cite{KNLSYSTEM:HOME} is composed of 
{8208} nodes connected by Intel Omni-Path network
with full bisectional bandwidth of 12.5 GB/s. 
Each node has an Intel Xeon Phi 7250 chip with 68 cores running at 1.4 GHz and 16 GB MCDRAM, 
and 96 GB DDR4 Memory. 
The theoretical peak performance of the node is 
3.0464 TFLOPS in double precision, resulting 
\num{25.004} PFLOPS for the whole system.

The KNL node has several clustering modes;\textit{All--to--All}, \textit{Quadrant/Hemisphere}, 
and \textit{SNC-4/SNC-2} for the core usage,
and memory modes; \textit{flat} and \textit{cache} for the MCDRAM. 
We use Quadrant-cache mode in this work.

\subsection{Weak-scaling up to 128 MPI processes}
We first present the weak-scaling up to 128 MPI processes varying the local lattice 
size $N_{LS}^3\times N_{LT}$ and the number of MPI-RANKs, whose four-dimensional sizes $P_X\times P_Y\times P_Z \times P_T$,
 are as follows:
$1\times 1\times 1\times 1 =    1$,
$2\times 1\times 1\times 1 =    2$,
$2\times 2\times 1\times 1 =    4$,
$2\times 2\times 2\times 1 =    8$,
$2\times 2\times 2\times 2 =   16$,
$4\times 2\times 2\times 1 =   16$,
$4\times 4\times 2\times 1 =   32$,
$4\times 4\times 4\times 1 =   64$,
$4\times 4\times 4\times 2 =  128$.

To benchmark the program 
we fix the number of multiplication $\hat{D}_{ee}$ in the nested-BiCGStab algorithm at {4000}, 
where every 500th multiplication is an outer-multiplication and the others are inner-multiplications.
We measure the performance (timing and FLOPS) of the inner-BiCGStab (single precision).
Two MPI-RANKs are placed on each node except for a single process execution.

Figure~\ref{fig:SOLVE_TIME_SMALL_WS} shows the time and the performance per MPI-RANK of 
the inner solver as defined from the line 3 to line 18 in Alg.~\ref{alg:InnerloopSP}, for which we define a notation of \textbf{SOLVE}.
At 16 MPI-RANKs, there are two configurations on the parallelism; 
$2\times 2\times 2\times 2 = 16$ and $4\times 2\times 2\times 1 = 16$, to compare the performances 
in three-dimensional and four-dimensional MPI partitioning.
The cases with $P_T=2$ (symbols showing lower performance at MPI-RANKs = 16 and 128) have 
a penalty due to the parallelization overhead caused by the SIMD vectorization in the temporal direction.

The plateau for the cases with $24^3\times N_{LS}$ (middle panels in Figure~\ref{fig:SOLVE_TIME_SMALL_WS})
starts from MPI-RANKs $= 2$ or $=4$, 
while it starts from MPI-RANKs $= 8$ with the cases with $16^3\times N_{LS}$ (left panels in Figure~\ref{fig:SOLVE_TIME_SMALL_WS}).
Although we are not fully understood the performance fluctuation seen on the larger local lattice sizes (right panels),
it could be caused by the huge memory consumption or memory thrashing.

Figure~\ref{fig:MULT_PERF_SMALL_WS} shows the performance of \textbf{MULT} with and without the communication time
to see the effect of the communication-computation overlapping.
The communication-computation overlapping works for the local lattice sizes larger than $24^3\times N_{LS}$,
because the performance degradation from upper to lower figures is invisible on the middle and right panels.
This is also confirmed by the timing measurement for the completion of receiving data in \textbf{MULT\_PST}.
The local lattice sizes larger than $24^3\times 24$ are sufficient to achieve a good weak-scaling properties.

\begin{table*}[tbh]
\centering
\caption{Timing and performance per node}
\label{tab:perfWS}
\begin{tabular}{ll|lllll}\hline
Lattice size& \# of nodes & \textbf{SOLVE} & \textbf{MULT}(w/o comm.) &  \textbf{MULT}(w/ comm.) & \textbf{YCOPY} & \textbf{MPI\_Allreduce} \\
\hline\hline
 $200^3 \times              800$  & 1000 & 90.109 [\si{s}] & 50.874 [\si{s}] & 51.002 [\si{s}] & 0.031 [\si{s}] & 14.627 [\si{s}] \\
 $400   \times 200^2 \times 800$  & 2000 & 92.340          & 51.067          & 51.187          & 0.025          & 16.531 \\
 $400^2 \times 200   \times 800$  & 4000 & 94.481          & 50.991          & 51.145          & 0.057          & 18.887 \\
 $400^3 \times              800$  & 8000 & 98.794          & 50.420          & 50.543          & 0.031          & 23.554 \\
\hline\hline
\multirow{2}{*}{Lattice size}& 
\multirow{2}{*}{\# of nodes} &
 \textbf{SOLVE}& \textbf{MULT}(w/o comm.) & \textbf{MULT}(w/ comm.) &
\multirow{2}{*}{-} & 
\multirow{2}{*}{-} \\
 & & [GFLOPS/node] & [GFLOPS/node] & [GFLOPS/node] & & \\
\hline\hline
 $200^3 \times              800$  & 1000 &  356.2 &  561.6 &  560.2 & & \\
 $400   \times 200^2 \times 800$  & 2000 &  347.6 &  559.5 &  558.1 & & \\
 $400^2 \times 200   \times 800$  & 4000 &  339.7 &  560.3 &  558.6 & & \\
 $400^3 \times              800$  & 8000 &  324.9 &  566.6 &  565.2 & & \\
\hline
\end{tabular}
\end{table*}

Let us look into the performance of \textbf{MULT} in details in terms of the arithmetic intensity and off-chip memory bandwidth.
For the single node execution, it achieves {400--470 [GFLOPS/RANK]} using 64 OpenMP threads per RANK 
on sufficiently large local lattice sizes.
The computation kernel \textbf{MULT} requires 180 complex numbers for {2232} FLOP per site, 
which corresponds to 
{0.645 Byte/FLOP} for the arithmetic intensity in single precision. 
The quark field at a site can be reused by eight times and the gluon field at the link by two times.
The arithmetic intensity can be reduced to 
{0.301 Byte/FLOP} in the ideal case that all data-access hit the cash memory.

The Intel Xeon Phi 7250 chip has 6.0928 TFLOPS for the peak performance in single precision,  
     115.2 GB/s for DDR4 memory bandwidth, and 
  475--490 GB/s~\cite{intelmeasure} for the MCDRAM bandwidth.
From these memory bandwidth, the kernel performance in the all-cache-hit case is expected to be:
\begin{itemize}
\item[1)]        383 GFLOPS with the DDR4 memory bandwidth,
\item[2)] 1578--1628 GFLOPS with the MCDRAM bandwidth, 
\end{itemize}
and the performance in the all-cache-miss case decreases to:
\begin{itemize}
\item[3)]        179 GFLOPS with the DDR4,
\item[4)]   736--760 GFLOPS with the MCDRAM. 
\end{itemize}
Our best performance observed at the single rank case is 
better than those [1) and 3)] with the DDR4 memory and 
 worse than those [2) and 4)] with the MCDRAM.
It could be the reason for the bad performance compared with the theoretical performance with MCDRAM 
that the single rank execution launched with 64-thread OpenMP consists of 2 SMT on 32 cores could 
not extract the best MCDRAM bandwidth.

In multiple MPI-RANK cases the \textbf{MULT} performance reaches {280--310 GFLOPS/MPI-RANK. }
In this case two MPI-RANKs are launched on a node and the single node performance becomes {560--620 GFLOPS/node.}
Though this performance number includes the overhead from 
splitting the kernel for the communication-computation overlapping,
the number approaches to the theoretical performance of 4).
The efficiency of the kernel performance is about {9--10\%} in 
multiple MPI-RANK cases with three-dimensional parallel partitioning.

\subsection{Weak-scaling towards 16000 MPI processes}

We benchmark the weak-scaling properties of the program using  8000 nodes of the OFP system.
We have fixed the number of multiplication $\hat{D}_{ee}$ in the nested-BiCGStab algorithm at ${2000}$ 
consisting of 4 sets of 500 inner-multiplications, where each set of inner-multiplications is 
followed by one outer-multiplication. 
We vary the number of MPI-RANKs from $10^3\times 2 =$ {2000} to $20^3\times 2 =$ \num{16000}
with a local lattice size fixed at $N_{LS}^3\times N_{LT}= 20^3\times 400$.

Table~\ref{tab:perfWS} shows the result of the weak-scaling benchmarking. 
The upper rows show the timing and the lower ones the performance per node.
\textbf{SOLVE} shows the time spent in the inner-BiCGStab Alg.~\ref{alg:InnerloopSP}.
\textbf{YCOPY} shows the time not hidden by the communication-computation overlapping 
for the nearest-neighbor communication in \textbf{MULT}. 
\textbf{MULT}(w/o comm.) shows the timing without the nearest-neighbor communication and
\textbf{MULT}(w/ comm.) includes the communication time \textbf{YCOPY}.
\textbf{MPI\_Allreduce} shows the time consumed in calling \textbf{MPI\_Allreduce} in
the reduction operations for the dot-product and the vector norm. 
The time of \textbf{MPI\_Allreduce} involves the timing comes from the barrier synchronization.
In Table~\ref{tab:perfWS} we observe that the weak-scaling performance of \textbf{MULT} is almost ideal, while 
the entire \textbf{SOLVE} performance shows a little degradation caused primarily by the \textbf{MPI\_Allreduce}.
{The performance of \textbf{SOLVE} reaches 2.6 PFLOPS sustained using 8000 nodes of the OFP system.}

\subsection{Realistic Simulation}

Having observed a good performance in the weak-scaling test on a $400^3\times 800$ lattice,
we executed a realistic simulation on the lattice with gauge field configurations 
at the coupling parameter $\beta=7.90$,
which is expected to have the lattice cut-off of $1/a\sim 25$ [\si{GeV}],
with the Wilson plaquette gauge action.
The box size becomes $aN_{X,Y,Z}=3.152$ [\si{fm}] which is sufficiently large to put a proton in the box.

Figure~\ref{fig:reshist} shows the residual history with $\kappa=0.13310$ and $c_{\mathrm{SW}}=1.30255$ 
on a gauge field, where the inner-solver (SP solver) is called four times from the outer-solver (DP solver) 
to obtain the final solution with double precision.
The timing to obtain the solution in double precision was 225.2 [\si{s}] in which 
\textbf{SOLVE} (SP Solver) timing was 205.0 [\si{s}] running at 2.57 [PFLOPS].

\renewcommand{\figscale}{0.45}
\begin{figure}[t]
    \centering
    \includegraphics[clip,scale=\figscale,trim=0 1 0 2]{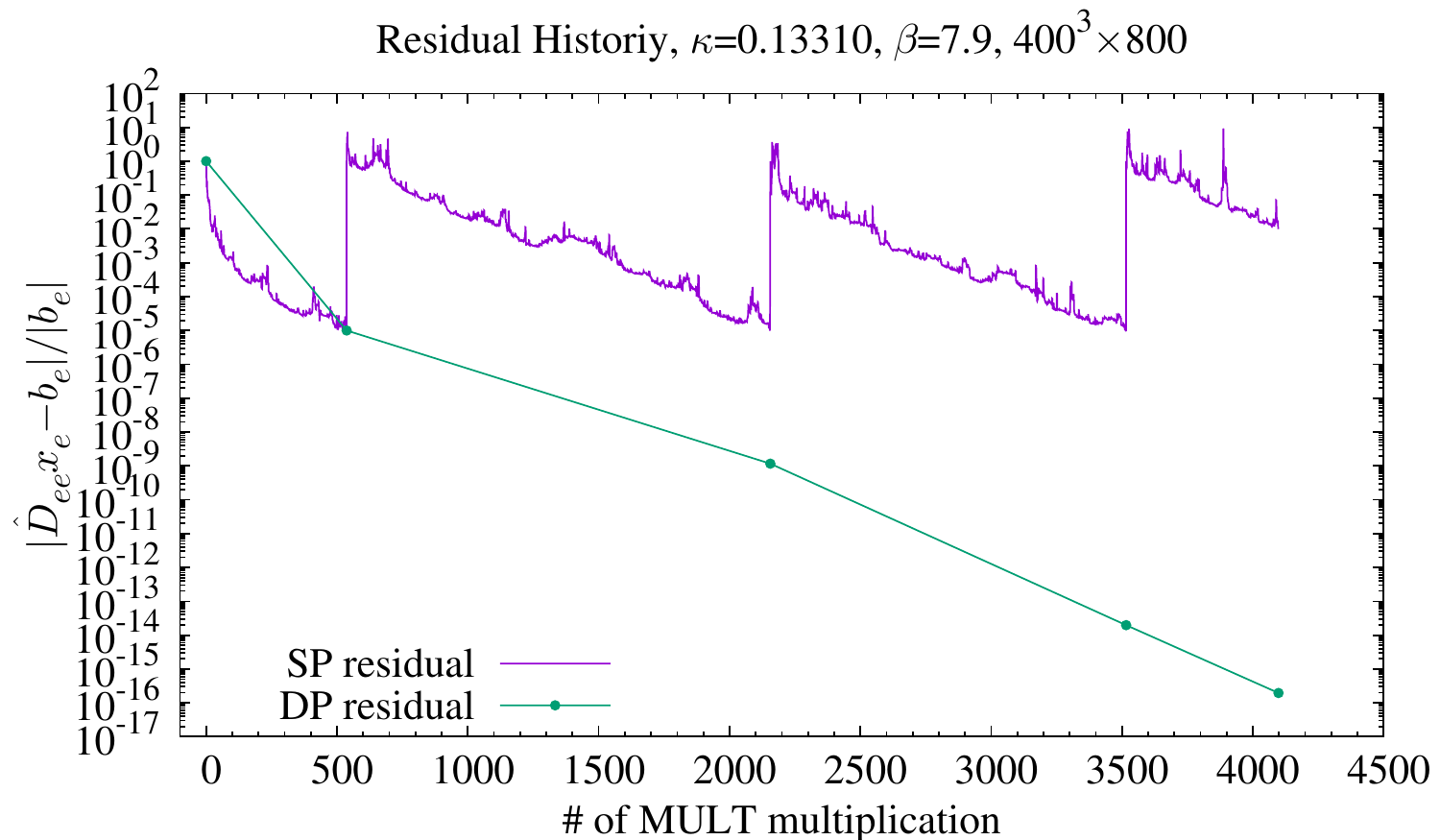}
    \caption{Solver residual history at $\kappa=0.13310$ and $\beta=7.9$ on a $400^3\times 800$ lattice.}
    \label{fig:reshist}
\end{figure}

\section{Conclusions}
\label{sec:last}

We have developed and optimized the quark solver for the Oakforest-PACS system.
We employed the nested-BiCGStab algorithm and the mixed-precision technique 
to utilize the high performance of single precision arithmetic of the system.
Basic performance tunings including
vectorization with the SIMD intrinsic functions, loop tiling with OpenMP threading, and prefetching, were applied.
To reduce the communication overhead, we implemented the communication-computation overlapping using MPI-offloading technique.

The benchmarking results up to 128 MPI processes showed a good weak-scaling property for local lattice sizes 
larger than $24^4$ indicating the communication overhead was hidden by the computation 
with our implementation. 
We also examined the large scale benchmarking up to 16000 MPI processes. 
As we did not implemented communication-avoiding algorithm, \textbf{MPI\_Allreduce} including collective synchronization timing
was a deficit in achieving better peak performance.
However we observed that the performance scaled well at the local lattice size of $20^3\times 400$
towards \num{16000} MPI processes.
On the lattice size of $400^3\times 800$, we observed 2.6 [PFLOPS] sustained using \num{16000} MPI processes with
 8000 nodes of the system.

\section*{Acknowledgments}
The authors would like to thank JCAHPC for their technical support and CPU hours of 
the Oakforest-PACS system for the trial use during December 2016 - March 2017. 
This work was supported in part by the Intel Parallel Computing Center 
at the Center for Computational Sciences (CCS), University of Tsukuba.

\end{document}